\documentclass[pra,aps,superscriptaddress,balancelastpage,onecolumn]{revtex4}
\usepackage{amsmath}
\usepackage{graphicx}
\usepackage{amsfonts}
\usepackage{amssymb}
\usepackage{bbold}
\usepackage{eucal}
\usepackage{mathrsfs}
\usepackage{epsfig}
\usepackage{epstopdf}
\usepackage[normalem]{ulem}
\usepackage{color}

\providecommand{\U}[1]{\protect\rule{.1in}{.1in}}
\providecommand{\U}[1]{\protect\rule{.1in}{.1in}}

\begin{document}

\title{Perfect Transfer of Entanglement and One-Way Quantum Steering via Parametric Frequency Converter in a Two-mode Cavity Magnomechanical System}
\author{Amjad Sohail}
\email{amjadsohail@gcuf.edu.pk,amjadss@unicamp.br}
\affiliation{Department of Physics, Government College University, Allama Iqbal Road,
Faisalabad 38000, Pakistan}
\affiliation{Instituto de Física Gleb Wataghin, Universidade Estadual de Campinas, Campinas, SP, Brazil}
\author{Allah Nawaz}
\affiliation{Department of Physics, Government College University, Allama Iqbal Road,
Faisalabad 38000, Pakistan}
\author{Hazrat Ali}
\affiliation{Department Physics, Abbottabad University of Science and Technology, Havellian, 22500, KPK, Pakistan}
\author{Rizwan Ahmed}
\email{rizwanphys@gmail.com}
\affiliation{Physics Division, Pakistan Institute of Nuclear Science and Technology
(PINSTECH), P. O. Nilore, Islamabad 45650, Pakistan.}
\author{Marcos César de Oliveira}
\email{marcos@ifi.unicamp.br}
\affiliation{Instituto de Física Gleb Wataghin, Universidade Estadual de Campinas, Campinas, SP, Brazil}
\begin{abstract}
We study the effects of a parametric frequency converter in a two-mode cavity system where one of the cavity mode is coupled with yttrium iron garnet (YIG) via magnetic dipole interaction. Parametric frequency converter acts as a nonlinear source for enhanced entanglement among all bipartitions and asymmetrical quantum steering. The behavior of the two types of quantum correlations
are shown to be dependent on parametric coupling and the associated phase factor $\phi$. We show that cavity-cavity entanglement and cavity-phonon entanglement (cavity-magnon entanglement) decreases (increases) with the increase of the parametric phase factor $\phi$.
In addition, generated entanglements  in the present system have shown to be more robust against the thermal effects, with the inclusion of the parametric converter as compared with the bare cavity case.
Another intriguing finding is the asymmetric one-way steering, where we notice that magnon and phonon modes can steer the indirectly coupled cavity modes, yet the steering in swapped direction is not observed. It is of great interest that the perfect transfer of entanglement and quantum steering is achieved among different modes by adjusting the system's parameters. In fact, our protocol for these transferring processes suggests a different approach to the processing and storage of quantum information.
\end{abstract}

\maketitle
\section{Introduction}

In recent years, considerable attention has been devoted to quantum mechanical systems aimed at achieving coupling between phonons and magnons through magnetostrictive interactions.\cite%
{magnom1,magnom2,magnosystem,magnosystem2,OP1,OP2}. Presently, research has been focused in the use of interactions, such as radiation pressure, Coulomb or piezoelectric interactions present in conventional opto-mechanical system\cite%
{rmp,optom1,optom2,ent,ent3,elect,elect2,piezo}. The study of cavity-magnon interactions, enabled by the introduction of yttrium-iron-garnet (YIG) spheres, has advanced magnetostrictive interactions within cavities. In these magnomechanical systems, the YIG sphere undergoes geometric squeezing \cite{ent3}, driven by shifts in magnetism caused by magnon excitations. This development has laid the foundation for an entirely new field of quantum magnomechanics. From the perspective of cavity magnomechanics, ferrimagnetic systems like the YIG sphere placed inside a microwave cavity, have
gained a great interest, recently. It has been noted that the YIG sphere can achieve strong coupling with the microwave photons in a high-quality cavity, leading to cavity magnons. As a result, many concepts that were first introduced in cavity magnomechanics\cite{magnom1}. 
Other fascinating developments in the context of cavity magnon systems include the coupling of a single superconducting qubit to the Kittel mode and the observation of bistability \cite{2016}.

The potential applications of magnetostrictive interactions in quantum information processing have been largely overlooked for a long time, due to their negligibly small magnitude in metallic and dielectric materials \cite{2016}. However, novel magnetic materials present a range of actual possibilities, where magnetostrictive interactions have gained prominence, enabling more tunable and versatile applications in information processing. Recently, such systems have been employed in ultrafast optics, achieving functionality even on the picosecond timescale \cite{25,26,27}. Many other interesting
phenomena have also been investigated as a result of intense magnetostrictive interaction, such as the interaction of a single superconducting qubit with a magnon mode or a three-level atom with a magnon \cite{MAOP}. Moreover, a large number of researchers have employed magnetomechanical systems \cite{QIP,QIP2,QIP3} to study important quantum information processing protocols. These include mechanical excitation \cite{QIP4} and quantum entanglement \cite{QIP5} methods for storage of optical information. In addition, numerous studies have been conducted exploring the unique characteristics of magnetostrictive interactions \cite{OP1,OP2,OP3}.

During the last two decades, many research groups have been involved in preparation of optomechanical/magnomechanical entanglement. These systems, inherently nonlinear in nature, have been the subject of extensive investigation. Furthermore, considerable attention has been given to medium assisted optomechanical/magnomechanical systems e.g., cavity having optical Kerr-type non-linearity \cite{OP4,MYJS}, atomic ensembles \cite{OP5,OP6,OP7}, optical parametric amplifier(OPA) \cite{bakht},
etc. It has been observed that quantum entanglement would be enhanced, as  the  presence of a nonlinear medium in anoptomechanical cavity increases the coupling between a movable mirror and a cavity mode, which in
turn leads to the increased cavity photon number \cite{OP9}, enhancing the incident radiation pressure upon the mechanical resonator (mirror). This results in various enhanced quantum features, making such medium-assisted systems pertinent.

As mentioned earlier, medium-assisted cavity optomechanics has enhanced many of quantum effects, as compared with the bare cavity optomechanical setups. One such optical element, that can be used as a medium inside an optomechanical cavity is the optical parametric converter (OPC). Parametric frequency converter as an entangler has been discussed in detail by Lee and collaborators in the past \cite{IQA1}. It is worth noting that, sum-frequency generation is a nonlinear optical process and it could be used in frequency conversion of an entangled state, as shown in a study by Tansilli et al. in \cite{IQA2} and by Tan et al. in \cite{IQA3}. Here we would like to point out a seminal study by Bergeal et al. \cite{PRLPC1}. They have observed the correlation between two modes of radiation in microwave (MW) regime through the quantum noise amplification via Josephson parametric converter. They measured the correlation by an interferometric experimental setup. In another interesting study, Abdo et al. demonstrated full frequency conversion in MW regime \cite{PRLPC2}. They used a Josephson device for three wave mixing being pumped at difference frequencies of fundamental eigenmodes. In the light of above experimental studies, it is feasible to employ parametric conversion process in MW regime. 

It is worthy to mention that there are many other parametric processes e.g., parametric amplification(OPA) and Non-degenerate (NOPA) \cite{IQA4} have been studied for the generation on entanglement, non-classical states of cavity field etc. \cite{IQA5}. In any type of parametric optical process, various factors are involved in the system dynamics. These factors could be controlled, in order to obtain the optimal quantum effects at mesoscopic scales. Hussain et. al. has shown that by employing degenerate optical parametric amplifier in a magnomecahnical system leads to robust and enhanced bipartite and tripartite entanglement \cite{bakht}.

In the present study, we consider a two-mode cavity system. The two modes are connected via a parametric frequency converter which acts as an entangler. Furthermore, one of the cavity modes is coupled with a YIG sphere via a magnetic dipole interaction. However, phonon mode (vibrational mode) is coupled to the magnon mode via the magnetostrictive effect. We have investigated the entanglement and quantum steering for various system parameters and found that this enhancement depends strongly upon the parametric gain and the  parametric phase of the parametric frequency converter. In addition, bipartitions present in the system demonstrate a complementary
relation, implying that the entanglements 
can be entirely/partially transferred to each other by tuning the parametric gain. Furthermore, cavity-magnon and cavity-phonon entanglements are found to be robust against thermal noise effects than cavity-cavity entanglement.

The paper is organized as follows. We discuss the theoretical model and the pertinent Hamiltonian in Section II and the dynamical equations are described in Sec. III. We define the quantum correlations like entanglement and quantum steering with their quantification measures in Sec. IV. Discussions about the obtained results are presented in section V. Finally, a conclusion encloses the paper in Sec. VI.
\begin{figure*}[tbp]
\centering
\includegraphics[width=0.95\columnwidth,height=3.6in]{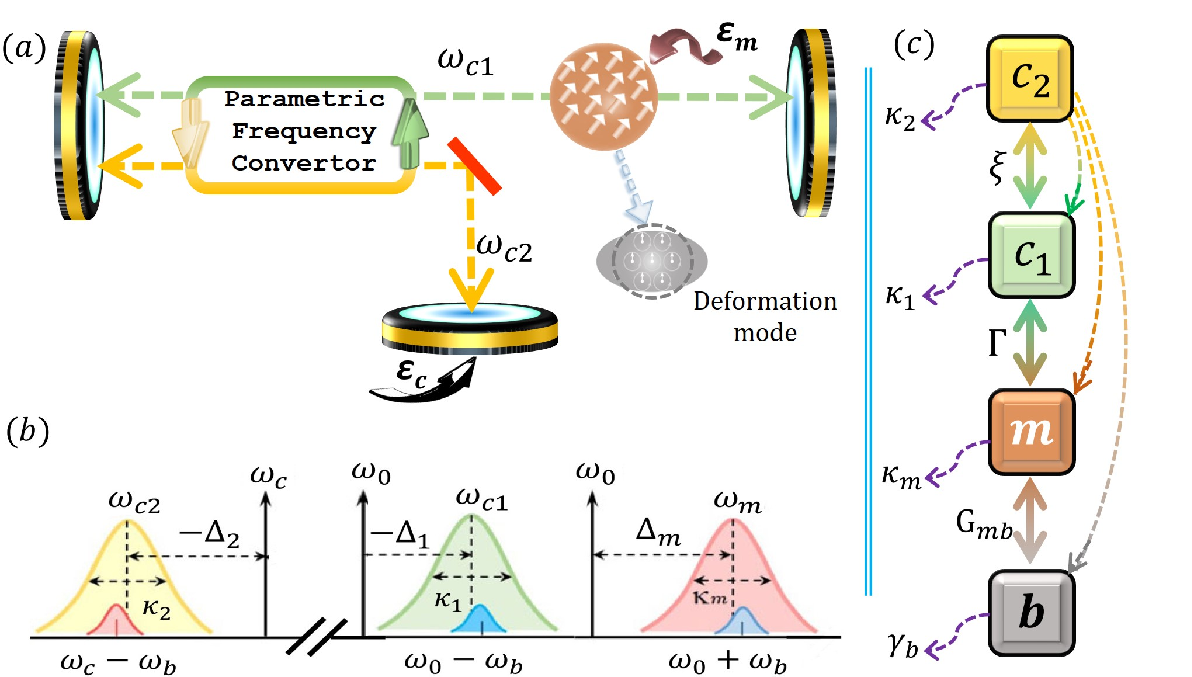} \centering
\caption{(a) Schematic illustration of our proposed system where the two cavity modes are coupled via parametric frequency converter. A YIG sphere
couple with MW cavity modes-1 through magnetic dipole interaction. In
addition, vibrational motion is considered as a phononic mode, which couples with the
magnonic mode owing to magnetostrictive effect. A microwave field with frequency $\omega_{0}$ and amplitued $\varepsilon_{m}$ drives the magnon while an external laser field with frequency $\omega_{c}$ and amplitued $\varepsilon_{c}$ drives the cavity. (b) Linewidths and Mode frequencies adopted in the protocol. When the cavity mode-1 and magnon mode match the red (Stokes) at $\omega_{0}-\omega_{b}$ and blue (anti-Stokes) sidebands at $\omega_{0}+\omega_{b}$, respectively, and moreover, when  the cavity mode-2 is resonant with red (Stokes) sideband with $\omega_{c}-\omega_{b}$ of the external laser drive field, stationary entanglement between cavity mode-2 with three other modes is established. (c) The coupling configuration and the corresponding decay rates are presented.}
\end{figure*}
\section{THEORETICAL MODEL AND HAMILTONIAN OF THE SYSTEM}
Consider a magnomechanical system having a parametric converter inside the cavity. The cavity contains two microwave modes, a magnonic  and a phononic modes as shown in Fig.1. The magnons are originated from  collective excitation of spins inside a ferrimagnet YIG sphere. The YIG sphere is exposed to a uniform bias magnetic field in the $z$-direction, which causes the magnon modes to get excited. These modes then couple with the cavity photons via magnetic-dipole interactions. 
Moreover, the magnetostrictive force, which deforms the sphere's geometrical structure and creates the magnon-phonon coupling, causes vibrations in the YIG sphere, which is a representation of the mechanical mode \cite{mag3,Kittel,Zha}. In general, the magnetostrictive interaction is a weak  interaction. Dependent upon the resonance frequencies of the magnon and phonon modes, however, it can become strong when the magnon frequency considered to be much stronger than the mechanical frequency and this becomes possible if we drive the YIG sphere with a strong microwave field. The total Hamiltonian of describing the system reads
\begin{eqnarray}
H/\hbar &=&\sum_{j=1}^{2}\Delta _{j}c_{j}c_{j}^{\dag }+\Delta
_{m_{0}}m^{\dag }m+\frac{\omega _{b}}{2}\left( q^{2}+p^{2}\right) +\Gamma
\left( c_{1}m^{\dag }+c_{1}^{\dag }m\right) +\mathcal{G}_{mb}m^{\dag }mq
\notag \\
&&+\xi\left[c_{1}^{\dag }c_{2}\exp (i\phi )+c_{2}^{\dag }c_{1}\exp (-i\phi
)\right]+i\varepsilon
_{m}\left( m^{\dag }-m\right) +i\varepsilon _{c}\left( c_{2}^{\dag }-c_{2}\right), \label{eqH}
\end{eqnarray}
where $\Delta _{m_{0}}=\omega _{m}-\omega _{0}$, $\Delta _{1}=\omega _{c1}-\omega _{0} $ and $\Delta
_{2}=\omega _{c2}-\omega _{c}$. Here, $\omega _{m}$ ($\omega
_{cj}$) is the resonance frequencies of the magnon ($jth$ cavity) mode while $%
\omega _{0}$ ($\omega _{c}$) is the frequency of the drive magnetic (laser) field. In first and
second term of the above Hamiltonian $c_{j} (c_{j}^{\dagger })$ and $m (m^{\dagger})$ represents the annihilation (creation) operator of the $j$th cavity mode
and magnon mode respectively. $\Delta _{m_{0}}$and $\Delta _{j}$ represents
detuning parameters associated with the respective modes. In the third term of eq. (\ref{eqH}), $q$
and $p$, (satisfying $[q,p]=i$), are the dimensionless position and momentum quadratures of
the mechanical mode and $\omega _{b}$ is the resonance frequency of the
phonon mode. The fourth term represents interaction between magnon mode and cavity-1 mode with optomagnonical coupling strength $\Gamma$
given by
\begin{equation*}
\Gamma =\nu \frac{c}{n_{r}}\sqrt{\frac{2}{\rho _{s}V_{yig}}},
\end{equation*}
where $V_{yig}$=$\frac{4\pi r^{3}}{3},\rho _{s},n_{r}$ and $\nu $
respectively are the volume, spin density, refractive index and Verdet
constant for the YIG sphere \cite{AORH}. Here we assume that there is a
strong coupling condition, where the optomagnonical coupling is larger
than the decay rates of the magnon and cavity modes \cite%
{kitt2,Tabuchi,kitt3,Goryachev} i.e. $\Gamma >c,\kappa _{m}$ and $\ \rho
_{s}=4.22\times 10^{27}m^{-3}.$ The fifth term represents the interaction
between magnon and phonon mode with the magnomechanical coupling strength $%
\mathcal{G}_{mb}$. 
Driving the magnon mode with a strong microwave field can improve the magnomechanical interaction, even though the single-magnon magnomechanical coupling is normally small.
The Rabi frequency $\Omega=\frac{5}{4}\gamma _{G} \sqrt{N}H_d$ indicates the strength of the coupling between the driving field of the microwave and the magnon, where $N=\rho_{s} V_{yig}$ stands for the YIG crystal's total spin number, and $H_d$ is the drive magnetic field's amplitude. 
The sixth term is parametric frequency converter (PFC) converter with $\xi$ and $\phi $ are the non linear gain and phase of the PFC. The last two terms are the input terms of cavity-2 $(c_{2})$ and magnon respectively. Finally, $\varepsilon _{c}=\sqrt{2\kappa\wp/\hbar\omega_{c}}$ is the coupling strength between the driving laser field and the cavity photon, where $\kappa$ is the decay rate of the cavity field and $P(\omega_{c})$ is the power (frequency) of the input laser field.

\section{The QUANTUM DYNAMICS}
Next, we discuss the quantum dynamics of two-cavity magnomechanical system by
employing the standard Langevin approach. By introducing the corresponding
effects of damping and noises, the following array of quantum Langevin
equations, describing the dynamics of the magnomechanical system, are as
given by
\begin{eqnarray}
\dot{q} &=&\omega _{b}p,  \label{LG} \\
\dot{p} &=&-\omega _{b}q-\mathcal{G}_{mb}m^{\dag }m-\gamma _{b}p+\zeta , \\
\dot{m} &=&-(\kappa _{m}+i\Delta _{m}^{0})m-i\Gamma c_{1}-i\mathcal{G}%
_{mb}mq +\varepsilon _{m}+\sqrt{2\kappa _{m}}m^{in}, \\
\dot{c}_{1} &=&-(\kappa _{1}+i\Delta _{1})c_{1}-i\Gamma m-i\xi c_{2}\exp
(i\phi )+\sqrt{2\kappa _{c}}c_{1}^{in}, \\
\dot{c}_{2} &=&-(\kappa _{2}+i\Delta _{2})c_{2}-i\xi c_{1}\exp (-i\phi
)+\varepsilon _{c}+\sqrt{2\kappa _{c}}c_{2}^{in},  \label{LEEE}
\end{eqnarray}%
where $\gamma _{b}$ is defined as the damping rate of mechanical mode, $%
k_{m} $ ($k_{j}$) denote the decay rate of the magnon mode ($j$th cavity
mode). $\zeta $, $m^{in}$ and $c_{j}^{in}$ are, respectively, the noise
operators for phonon mode, magnon mode and cavity mode. Noise operators for $%
j $th cavity and magnon mode must satisfy the correlation function \cite%
{Gardiner}
\begin{eqnarray}\left\langle c_{j}^{in\dagger }\left( t\right) c_{j}^{in}\left(
\acute{t}\right) \right\rangle &=&n_{j}\left( \omega _{j}\right) \delta \left(
t-\acute{t}\right) , \\ \left\langle c_{j}^{in}\left( t\right)
c_{j}^{in\dagger }\left( \acute{t}\right) \right\rangle &=&\left[ n_{k}\left(
\omega _{j}\right) +1\right] \left( t-\acute{t}\right) ,\\ \left\langle
m_{j}^{in\dagger }\left( t\right) m_{j}^{in}\left( \acute{t}\right)
\right\rangle &=&n_{m}\left( \omega _{m}\right) \delta \left( t-\acute{t}%
\right) , \\ \left\langle m_{j}^{in}\left( t\right) m_{j}^{in\dagger
}\left( \acute{t}\right) \right\rangle &=&\left[ n_{m}\left( \omega
_{m}\right) +1\right] \delta \left( t-\acute{t}\right) .\end{eqnarray} 
In addition, the
phonon damping rate satisfies the correlation function given by \begin{equation}
\langle \zeta ( \acute{t}) \zeta (t) + \zeta ( \acute{t}) \zeta (t)
\rangle/2=\gamma _{b}\left[ 2n_{b}\left( \omega _{b}\right) +1%
\right] \delta \left( t-\acute{t}\right).
\end{equation}
In all the above correlation
function, $n_{s}\left( \omega _{s}\right) =\left[ \frac{\exp \hbar \omega
_{s}}{k_{B}T}-1\right] ^{-1},s=j\left( j=1,2\right),b$ and $m$ are
the equilibrium mean thermal photon, phonon, magnon number respectively,
where $k_{B}$ denote the Boltzman constant and $T$ the temperature. 
It is important to mention here that the dynamical equation can be
linearized if the magnon mode is strongly driven such that, $\left\vert
\left\langle m\right\rangle \right\vert \gg 1$. Furthermore, the strong
interaction of the two MW cavity fields with the magnon mode leads to a large cavity
amplitude, i.e., $\left\vert \left\langle c\right\rangle \right\vert \gg 1$.
These two conditions allow us to linearize Eq.(\ref{LG}-\ref{LEEE}) 
by writing
each operator as a sum of steady-state (average) value and the fluctuation
part\cite{Shl1,Shl2}, i.e., $\mathcal{Q}=\left\langle \mathcal{Q}%
\right\rangle +\delta \mathcal{Q},(\mathcal{Q}=p,q,c_{k},m)$. 
Therefore, the steady-state averages of the operators are given by
\begin{eqnarray}
 p_{s} &=&0,\ \ \ \ \ \  \\
\ \ \ \ q_{s} &=&\frac{-g_{mb}}{\omega _{b}}\left\vert m_{s}\right\vert ^{2},
 \\
c_{2s} &=&\frac{\varepsilon _{c}-i\xi\exp (-i\phi )c_{1s}}{\kappa _{{2}%
}+i\Delta _{2}},  \label{EQ} \\
c_{1s} &=&\frac{-i\Gamma \left( \kappa _{{2}}+i\Delta _{2}\right)
m_{s}-i\xi\exp (i\phi )\varepsilon _{c}}{\left[ \left( \kappa _{{2}%
}+i\Delta _{2}\right) \left( \kappa _{{1}}+i\Delta _{1}\right) +G^{2}\right]
},  \label{CON}
\\
m_{s} &=&\frac{\varepsilon _{m}\Theta -\Gamma \xi\exp (i\phi )\varepsilon
_{c}}{\Theta \left( \kappa _{m}+i\Delta _{m}\right) +\Gamma ^{2}\left(
\kappa _{{2}}+i\Delta _{2}\right) } , \label{MAV}
\end{eqnarray}
where $\Theta =\left[ \left( \kappa _{{2}}+i\Delta _{2}\right) \left( \kappa
_{{1}}+i\Delta _{1}\right) +\xi^{2}\right] $ and $\Delta _{m}=\Delta _{m}^{0}+%
\mathcal{G}_{mb}\left\langle q\right\rangle $ is the effective magnon detuning which  incorporates a small frequency shift due to the magnomechanical
interaction. Also, $\Delta _{1}\Delta _{2}>>\kappa _{1}\kappa _{2}$, the steady state solution for the magnon takes a simpler form
\begin{equation}
m_{s}=\frac{\left[ \left( i\Delta _{2}\right) \left( i\Delta _{1}\right)
+\xi^{2}\right] \varepsilon _{m}-\Gamma \xi\exp (i\phi )\varepsilon _{c}}{%
\left[ \left( i\Delta _{2}\right) \left( i\Delta _{1}\right) +\xi^{2}\right]
\left( i\Delta _{m}\right) +\Gamma ^{2}\left( i\Delta _{2}\right) }.
\end{equation}
The linearized QLEs can be described by the set of quantum fluctuations
equations, given by
\begin{eqnarray}
\delta \dot{q} &=&\omega _{b}\delta p,  \label{Q} \\
\delta \dot{p} &=&-\omega _{b}\delta q-\mathcal{G}_{mb}\left( m_{s}\delta
m^{\dag }+m_{s}^{\ast }\delta m\right)-\gamma _{b}\delta p+\zeta ,  \label{P}
\\
\delta \dot{m} &=&-(\kappa _{m}+i\Delta _{m})\delta m-i\Gamma \delta c_{1}-i%
\mathcal{G}_{mb}m_{s}\delta q+\sqrt{2\kappa _{m}}\delta m^{in},  \label{M} \\
\delta \dot{c}_{1} &=&-(\kappa _{1}+i\Delta _{1})\delta c_{1}-i\Gamma \delta
m-i\xi\delta c_{2}\exp (i\phi )+\sqrt{2\kappa _{c}}\delta c_{1}^{in},
\label{C1} \\
\delta \dot{c}_{2} &=&-(\kappa _{2}+i\Delta _{2})\delta c_{2}-i\xi\delta
c_{1}\exp (-i\phi )+\sqrt{2\kappa _{c}}\delta c_{2}^{in},  \label{C2}
\end{eqnarray}
Now, the magnon and cavity fluctuation operators can be rewritten in terms of the quadratures $\delta x=\frac{1}{\sqrt{2}%
}\left( \delta m^{\dagger }+\delta m\right) $, $\delta y=\frac{i}{\sqrt{2}}%
\left( \delta m^{\dagger }-\delta m\right) $, $\delta X_{j}=\frac{1}{\sqrt{2}%
}\left( \delta c_{j}^{\dagger }+\delta c_{j}\right) $ and $\delta Y_{j}=%
\frac{i}{\sqrt{2}}\left( \delta c_{j}^{\dagger }-\delta c_{j}\right) $. By incorporating these quadratures into the above fluctuation equations, it can
be easy to write as a first-order 
differential equation in a compendious way as
\begin{equation}
\frac{d}{dt}{\digamma}(t)=\mathcal{M}\digamma (t)+\mathcal{N}(t),
\end{equation}
where $\mathcal{M}$ is the drift matrix, $\digamma (t)$ is the quantum
fluctuation and $\mathcal{N}(t)$ is the input noise vectors, which are all expressed
as
\begin{eqnarray}
\digamma (t) &=&[\delta q(t),\delta p(t),\delta x(t),\delta y(t),\delta
X_{1}(t),\delta Y_{1}(t),\delta X_{2}(t),\delta Y_{2}(t)]^{T},\\
\mathcal{N}(t)&=&[0,\zeta(t),\sqrt{2k_{m}}x_{1}^{in}(t),\sqrt{2k_{m}}y_{m}^{in}(t), \sqrt{2k_{1}}(X_{1}^{in}(t),Y_{1}^{in}(t)),\sqrt{2k_{2}}%
(X_{2}^{in}(t),Y_{2}^{in}(t))]^{T},\\
M&=&
\left(
\begin{array}{cccccccc}
0 & \omega _{b} & 0 & 0 & 0 & 0 & 0 & 0 \\
-\omega _{b} & -\gamma _{b} & -\mathcal{G}_{mb}\alpha & -\mathcal{G}%
_{mb}\beta & 0 & 0 & 0 & 0 \\
\mathcal{G}_{mb}\beta & 0 & -\kappa _{m} & \Delta _{m} & 0 & \Gamma & 0 & 0
\\
-\mathcal{G}_{mb}\alpha & 0 & -\Delta _{m} & -\kappa _{m} & -\Gamma & 0 & 0
& 0 \\
0 & 0 & 0 & \Gamma & -\kappa _{1} & \Delta _{1} & \mu & \nu \\
0 & 0 & -\Gamma & 0 & -\Delta _{1} & -\kappa _{1} & -\nu & \mu \\
0 & 0 & 0 & 0 & -\mu & \nu & -\kappa _{2} & \Delta _{2}
\\
0 & 0 & 0 & 0 & -\nu & -\mu & -\Delta _{2} & -\kappa _{2}%
\end{array}%
\right) ,
\end{eqnarray}
where $\alpha =\frac{m_{s}^{\ast }+m_{s}}{\sqrt{2}}$, $\beta =i\frac{%
m_{s}^{\ast }-m_{s}}{\sqrt{2}}$, $\mu=\xi\sin \phi$ and  $\nu=\xi\cos \phi$. If the magnon mode is driven by a strong field, one can observe that $\mathcal{G}_{mb}$ can easily be
enhanced (see Eq. (\ref{MAV})).
\section{QUANTUM CORRELATIONS}
We now discuss about the quantum correlations for various bipartitions.\\
\textbf{\textit{Stationary Entanglement}}:
Since, the drift matrix of our magnomechanical system is $8\times 8$ matrix,
the corresponding covariance matrix (CM) $V$ will also be a $8\times 8$
matrix, with the entries
\begin{equation}
V_{ij}(t)=\frac{1}{2}\left\langle \digamma _{j}(t^{\prime })\digamma
_{i}(t)+\digamma _{i}(t)\digamma _{j}(t^{\prime })\right\rangle .
\end{equation}
The CM of any system can be 
acquired by numerically solving steady-state Lyapunov equation
\cite{Parks,SA}
\begin{equation}
\mathcal{M}\mathcal{V}+\mathcal{V}\mathcal{M}^{T}=-\mathit{D},
\end{equation}%
where $\mathit{D}=$ is called the diffusion matrix whose matrix elements can be numerically obtained from the elements of the column vector
of the noise sources and utilizing the input noise correlations of the four modes of the system in $D_{ij}\delta (t-t^{\prime})=\frac{1}{2}\left\langle
\mathcal{N}_{i}(t)\mathcal{N}_{j}(t^{\prime})+\mathcal{N}_{j}(t^{%
\prime })\mathcal{N}_{i}(t)\right\rangle$.
As a result, we obtain matrix $\mathit{D}$, is given by
\begin{equation}
\mathit{D}= \mbox{diag}[0,\gamma _{b} \mathcal{T}_b,\kappa
_{m}\mathcal{T}_m ,\kappa _{m}\mathcal{T}_m,\kappa _{1}\mathcal{T}_1 ,\kappa _{1}\mathcal{T}_1,\kappa _{2}\mathcal{T}_2,\kappa _{2}\mathcal{T}_2],
\end{equation}
where $\mathcal{T}_s=\left( 2n_{s}+1\right), (s=b,m,1,2)$. To investigate the bipartite entanglement of the system, We employ the quantitative measure of the logarithmic negative by using the Simon
condition for Gaussian states \cite{Gonzalez,Vidal,Plenio,Adesso,MCO1}.
\begin{equation}
E_{N}=\max [0,-\ln 2(\nu ^{-})],
\end{equation}%
where $\nu ^{-}=$min eig$|\bigoplus_{j=1}^{2}(-\sigma _{y})\widetilde{%
\mathcal{V}_{4}}|$ defines the minimum symplectic eigenvalue of the CM which
is reduced to order of $4\times 4$. Here, $\widetilde{\mathcal{V}_{4}}%
=\varrho _{1|2}V_{in}\varrho _{1|2}$. Thus the uninteresting columns and rows
in\ $V_{4}$ can be removed to create a $4\times 4$ matrix of any
bipartition, which is represented by $V_{in}$. $\varrho _{1|2}$=diag$%
(1,-1,1,1) $=$\sigma _{z}\bigoplus \mathbb{1}$, represents a matrix that is
essentially used to carry out a partial transposition at the CM matrix
level. Here, $\mathbb{1}$ is the identity matrix and $\sigma $'s are the
pauli spin matrices. Moreover, $E_{N}>0$ establishes the existence of
entanglement between any bipartite entanglement in system.

\textbf{\textit{Quantum Steering}} Quantum steering, owing to its asymmetric features between any two interacting parties, differs significantly from entanglement, as it directly quantifies a one-way non-locality\cite{wise1, MCO2}. Based on quantum coherent information, and given the Gaussian state with two interacting modes, the quantum steerability in various directions is given by \cite{ZSSF,KILA}
\begin{figure}[tbp]
\begin{center}
\includegraphics[width=0.9\columnwidth,height=4.5in]{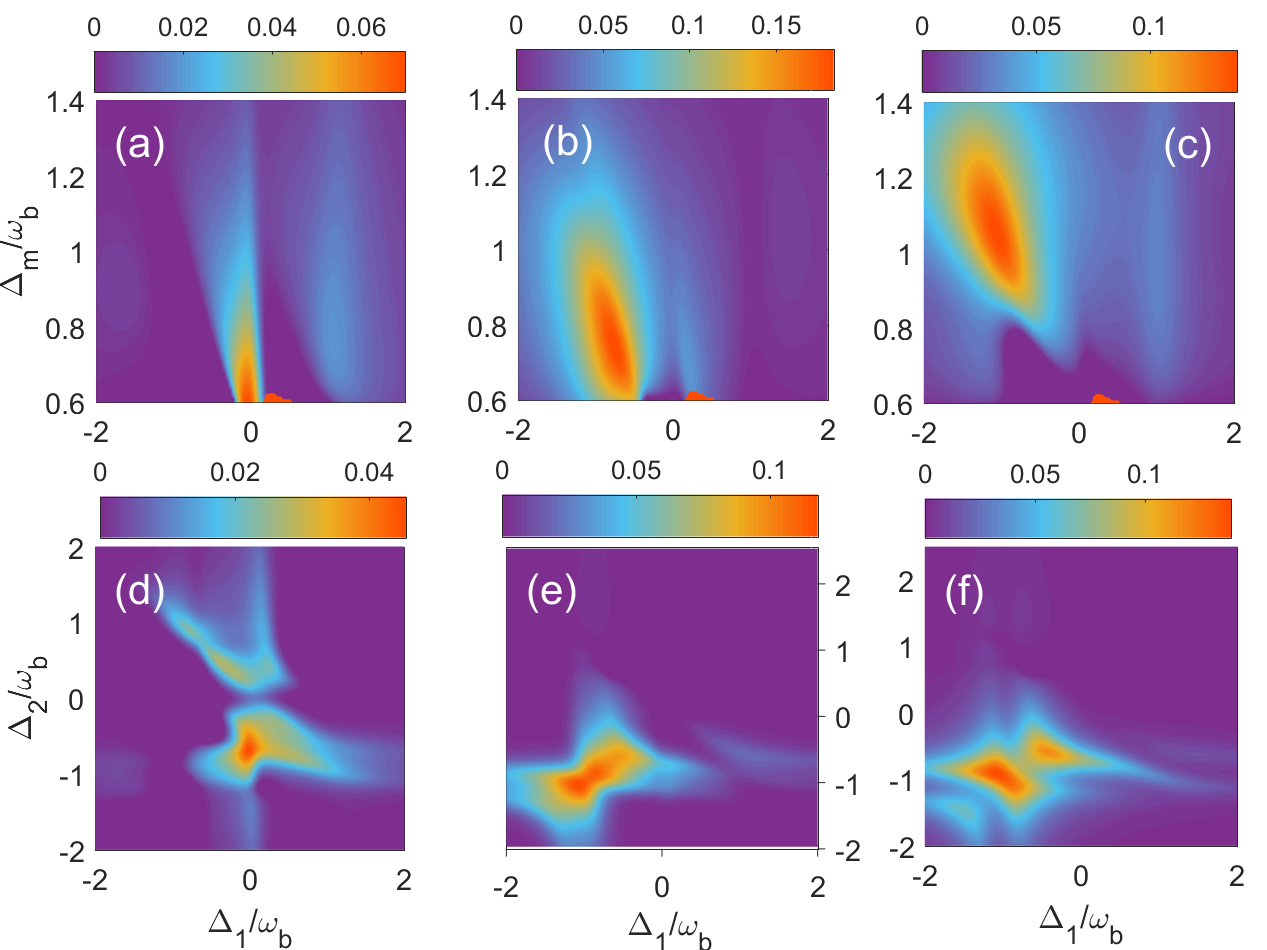}
\end{center}
\caption{Contour plot of bipartite entanglement (a)(d) $E^{N}_{c_{2}-c_{1}}$, (b)(e) $E^{N}_{c_{2}-m}$, and (c)(f) $E^{N}_{c_{2}-b}$ versus detunings $\Delta_{1}/\omega_{b}$ and (a-c) $\Delta_{m}/\omega_{b}$, when $\Delta_{2}/\omega_{b}=-1$ and (d-f) $\Delta_{2}/\omega_{b}$, when $\Delta_{m}/\omega_{b}=1$. Here we take $\xi=0.3 \omega_b$ and $\phi=0$. The additional parameters are detailed in the text.}
\end{figure}
\begin{eqnarray}
S_{\rho | \varrho } &=&\max \{0,\mathcal{R}(2\mathcal{V}%
_{u})-\mathcal{R}(2\mathcal{V}_{in})\},  \label{A} \\
S_{\varrho | \rho } &=&\max \{0,\mathcal{R}(2\mathcal{V}%
_{v})-\mathcal{R}(2\mathcal{V}_{in})\},  \label{B}
\end{eqnarray}%
where
\begin{equation}
\mathcal{V}_{in}=\left[
\begin{array}{cc}
\mathcal{V}_{u} & \mathcal{V}_{uv} \\
\mathcal{V}_{uv}^{T} & \mathcal{V}_{v}%
\end{array}%
\right], \label{C}
\end{equation}%
and $\mathcal{R}(\nu )=\frac{1}{2}\ln \det (\nu )$ is the R\'{e}nyi-2
entropy which quantify the degree of quantum steering. Each entry in Eq. (\ref{C}) is a $2\times 2$ matrix and the diagonal entries $\mathcal{V}_{u}$ and $\mathcal{V}_{v}$  stand for the modes u and v's reduced states, respectively.
The steerability in the direction
from mode $u$ to mode $v$ is characterized by $\mathcal{S}_{u|v}$ while from mode $v $
to mode $u$ defines the swapped direction, i.e.,  $S_{v | u }$. Furthermore, it is vital to understand steering effects i.e., $S_{u | v }=0$ and $S_{v | u }=0$ defines no-way
steering, $S_{u | v }>0$ and $S_{v |u }=0$ or $S_{u | v }=0 $ and $S_{v | u }>0$ characterizes one-way
steering and $S_{u | v}>0$ and $S_{v| u }>0$ shows two-way steering.
\section{RESULTS AND DISCUSSION}
We primarily address the key findings pertaining to entanglement in three distinct bipartitions, such as the entanglement between the 
two cavity modes $E^N_{c2-c1}$, cavity mode and the magnon mode $E^N_{c2-m}$, and cavity mode and the phonon mode $E^N_{c2-b}$. Entanglement between these modes is found to be generated by the entanglement transfer from the entanglement source. The types of the entanglement source, magnomechanical entanglement in the current system, are determined by the drives of magnon and cavity modes. Experimentally feasible parameters have been employed \cite{{2016}}, which are: $\omega_{c_{1}}=\omega_{c_{2}}=2\pi\times 10$ GHz, $\omega_{b}=2\pi\times10$ MHz, 
$\gamma_{b}=100$ MHz $\kappa_{{1}}= \kappa_{{2}}= 2\pi\times 1$ MHz and $\Gamma=2\pi\times 3.2$ MHz, $H_d=1.3\times10^-{4}, \gamma_{G}/2\pi=28$GHz/T, $r=250\mu$m, and $\wp=8.9$ mW.
Figures 2(a-c) are the contour plots representing the three bipartitions for a range of magnon and
cavity detunings for a fixed value of $\Delta_{2}=-\omega_{b}$, indicating that these bipartitions can be optimally entangled at various magnon and cavity detunings. It is easy to conclude that by manipulating the magnon and cavity detunings, the entanglement between various bipartitions can be redistributed. Next,  fig. 2(d-f) represents the same bipartitions against $\Delta_{1}$ and $\Delta_{2}$  for a fixed value of $\Delta_{m}=\omega_{b}$. It can be seen that the optimal value of $E^N_{c2-m}$  and $E^N_{c2-b}$ are obtained when both modes are in red-detuned region.
\begin{figure}[tbp]
\begin{center}
\includegraphics[width=0.9\columnwidth,height=4.5in]{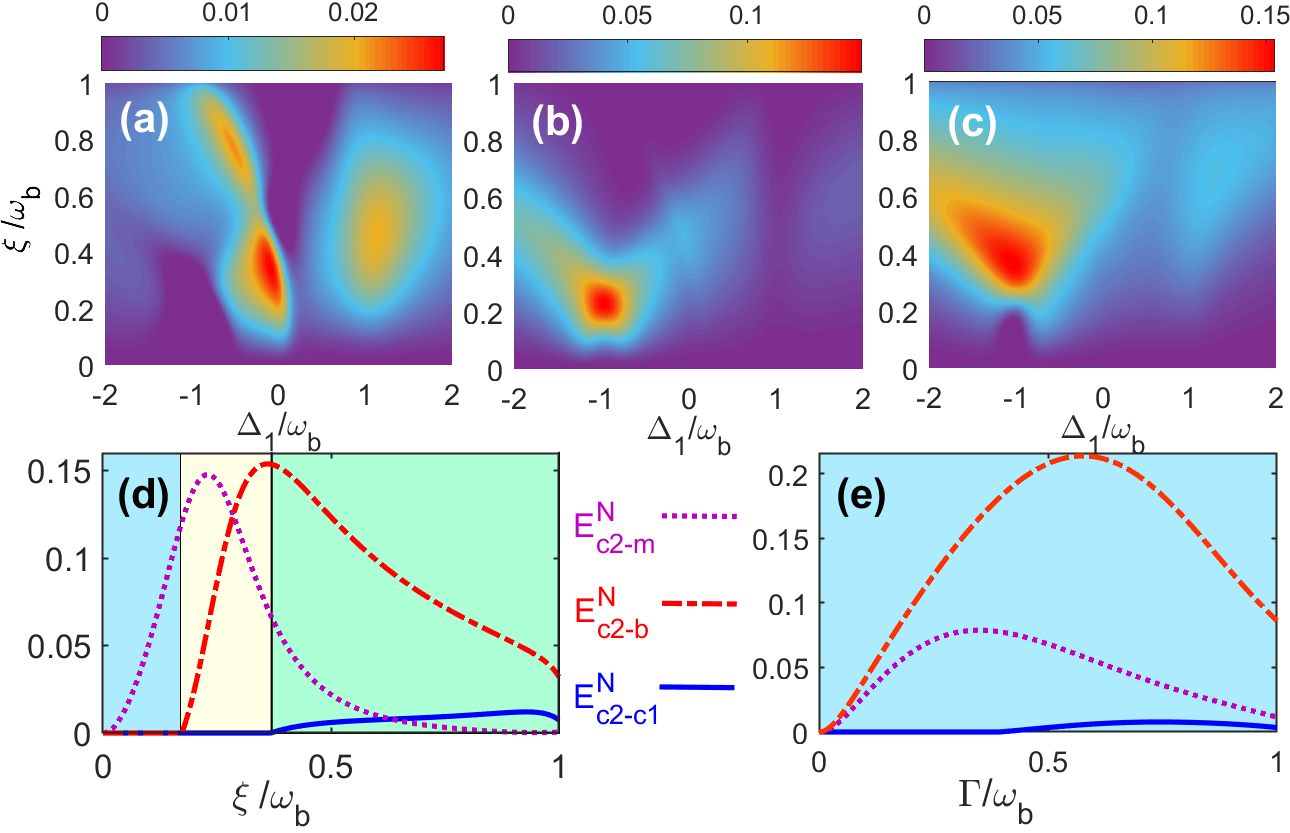}
\end{center}
\caption{Contour plot of bipartite entanglement (a) $E^{N}_{c_{2}-c_{1}}$, (b) $E^{N}_{c_{2}-m}$, and (c) $E^{N}_{c_{2}-b}$ versus detunings $\Delta_{1}/\omega_{b}$ for different value of $\xi$, keeping $\Delta_{2}=-\omega_{b}$, and $\Delta_{m}=\omega_{b}$. Plot of bipartite entanglement $E^{N}_{c_{2}-c_{1}}$ (solid blue line),  $E^{N}_{c_{2}-m}$ (dotted indigo line), and $E^{N}_{c_{2}-b}$ (dot-dashed red line) versus (d) $\xi$ and (e) $\Gamma$ when $\Delta_{1}=\Delta_{2}=-\omega_{b}$ and $\Delta_{m}=\omega_{b}$. We take $\Gamma=0.32\omega_{b}$ for (d) while $\xi=0.35\omega_{b}$ for (e). The rest of the parameters are given in the text.}
\end{figure}
\begin{figure}[tbp]
\begin{center}
\includegraphics[width=1\columnwidth,height=2.5in]{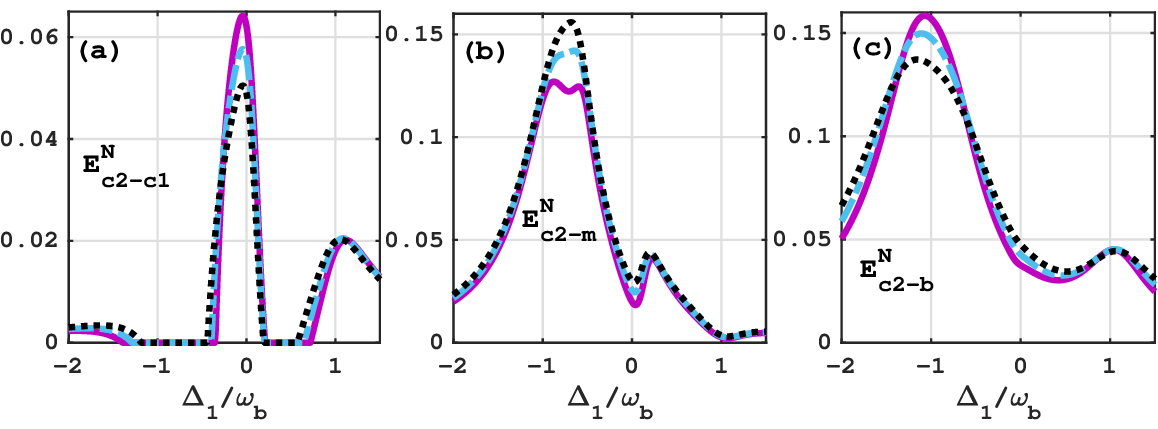}
\end{center}
\caption{Plot of bipartite entanglement (a) $E^{N}_{c_{2}-c_{1}}$, (b) $E^{N}_{c_{2}-m}$, and (c) $E^{N}_{c_{2}-b}$ versus detunings $\Delta_{1}/\omega_{b}$ for different value of $\varphi$, when $\varphi=0$ (indigo solid line), $\varphi=\pi/2$ (blue dot-dashed line), and $\varphi=\pi$ (black dotted line).  We choose the optimal values of the detunings from Fig. 2. The rest of the parameters are given in the text.}
\end{figure}
\begin{figure}[b!]
\begin{center}
\includegraphics[width=1\columnwidth,height=2.8in]{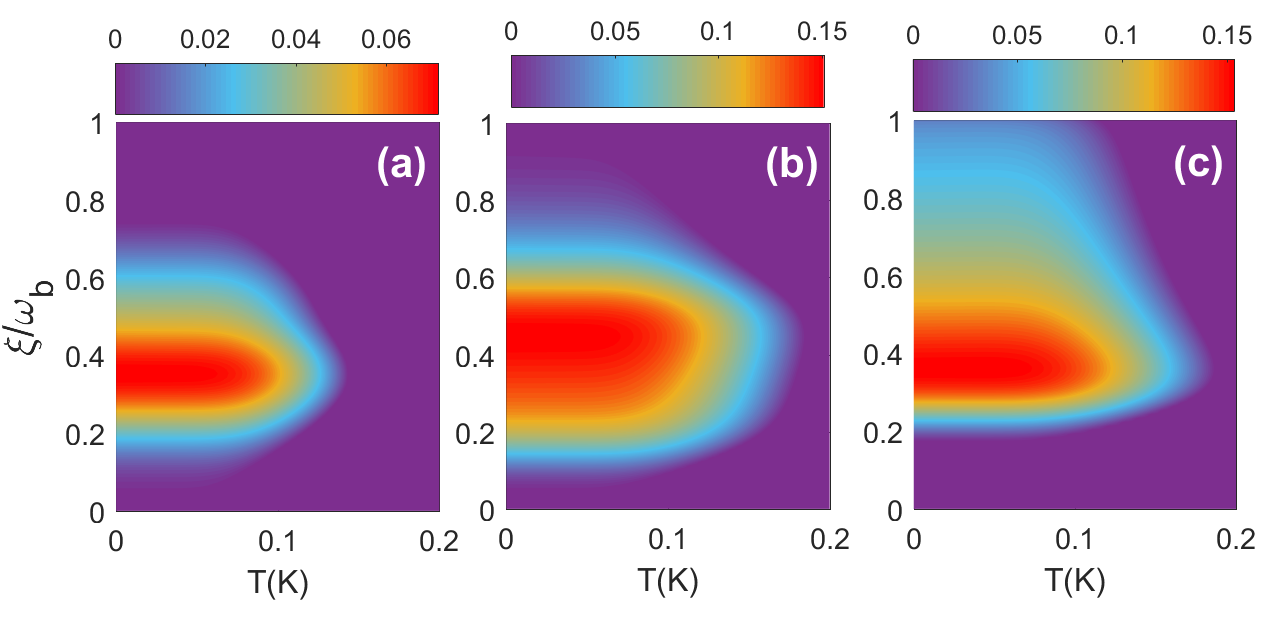}
\end{center}
\caption{Plot of bipartite entanglement (a) $E^{N}_{c_{2}-c_{1}}$, (b) $E^{N}_{c_{2}-m}$, and (c) $E^{N}_{c_{2}-b}$ versus temperature $T$(K) and $\xi$. We choose the optimal values from Fig. 2. The rest of the parameters are given in the text.}
\end{figure}

Next, fig, 3 (a-c) shows the effect of gain of the parametric converter on the three bipartitions, keeping fixed the value of $\Delta_{2}$, and $\Delta_{m}$. It can be seen that $E^N_{c2-c1}$  and $E^N_{c2-b}$ exist in both red and blue detuned region, however, $E^N_{c2-m}$ almost cease to exist in blue detuned region, On the other hand, the optimal value of $E^N_{c2-m}$ and $E^N_{c2-b}$ are obtained exactly at $\Delta_{1}=-\omega_{b}$, while $E^N_{c2-c1}$ is obtained when $\Delta_{1}=0$. As illustrated in fig. 3 (d), cavity-2-magnon entanglement $E^N_{c2-m}$ arises immediately after the gain of the parametric converter switched on and reaches the optimal value while rest of the two bipartitions are zero until $\xi=0.2\omega_{b}$. It is noteworthy that $E^N_{c2-b}$ can than be induced and while $E^N_{c2-m}$ tends to decrease when the PFC gain exceeds $\xi=0.2\omega_{b}$. Hence, one can safely say that $E^N_{c2-m}$ start transferring to $E^N_{c2-b}$ via the interaction $\propto mb^{\dagger}+m^{\dagger}b$. Furthermore, entanglements $E^N_{c2-m}$ and $E^N_{c2-b}$ manifestly show a decreasing trend while $E^N_{c2-c1}$ demonstrates the opposite upward tendency, with an increase in parametric gain $\xi$, which shows
the complete transfer of entanglement indirectly modes to directly coupled mode.
It is noteworthy to observe that both $E^N_{c2-c1}$ and $E^N_{c2-b}$ can be generate only when the parametric gain $\xi$ surpass  specific values. On the contrary, the entanglement $E^N_{c2-m}$ emerges immediately upon the introduction of the parametric gain $\xi$.
Furthermore, it is evident from fig. 3 (e) that entanglements $E^N_{c2-m}$ and $E^N_{c2-b}$ increase immediately as the microwave cavity-magnon coupling rate $\Gamma$ increases. $E^N_{c2-m}$ reaches its optimum value at
$\Gamma= 0.32\omega$. As $\Gamma$ continues to increase, $E^N_{c2-m}$ ($E^N_{c2-b}$) decrease (increase) which support our previous claim of shifting entanglement from $E^N_{c2-m}$ to $E^N_{c2-b}$) via $mb^{\dagger}+m^{\dagger}b$.
Additionally, the participation of cavity-cavity
entanglement (i.e. $E^N_{c2-c1}>0$) is observed for stronger optomagnonical coupling, when the rest two show decreasing trend.
A transfer between separate entanglements can be thought of as the result of this optomagnonical coupling strength modification, which weakens one entanglement while strengthening the other. As an important quantum phenomenon, entanglement transfer has a variety of physical applications in quantum computing \cite{ASTE}, quantum communication \cite{NZOU}, quantum sensing \cite{CLDF}, and quantum
key distribution \cite{TGVH}.
\begin{figure}[b!]
\begin{center}
\includegraphics[width=0.9\columnwidth,height=5.2in]{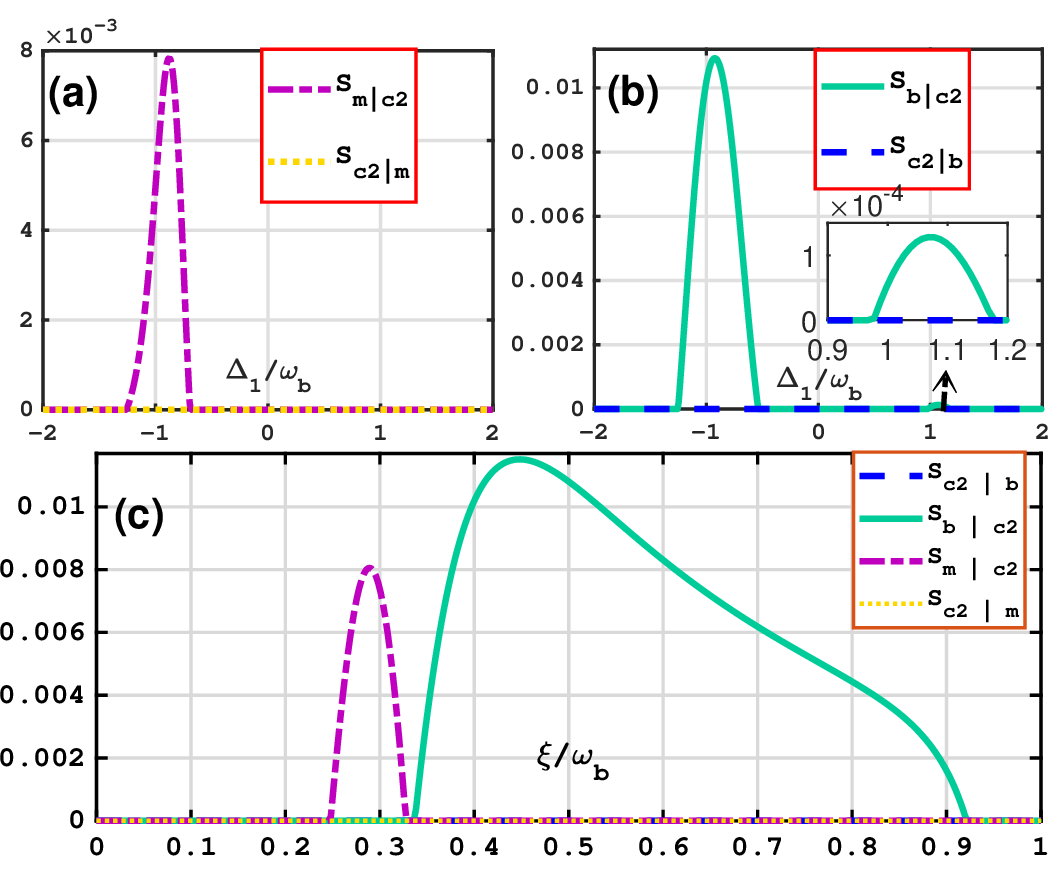}
\end{center}
\caption{Plot of steering (a) $S_{m | c2 } $ and $S_{c2 | m} $, (b) $S_{b | c2 } $ and $S_{c2 | b} $ versus normalized detuning $\Delta_{1}/\omega_{b}$. We choose the optimal values from Fig. 2. (c) Plot of steering $S_{m | c2 } $ and $S_{c2 | m} $, $S_{b | c2 } $ and $S_{c2 | b} $ versus normalized detuning $\xi/\omega_{b}$ when $\Delta_{1}=-\omega_{b}$. The rest of the parameters are given in the text.}
\end{figure}

In Fig. 4, we present all bipartite entanglements as a function of dimensionless detuning $\Delta_{1}/\omega_b$ for different values of the phase of the PFC. We observe that the phase of the PFC plays a vital role. The maximum enhancement of the entanglements $E^N_{c2-c1}$ and $E^N_{c2-b}$ is obtained for $\phi=0$ while the maximum enhancement of the entanglements $E^N_{c2-m}$ occurs when the phase of the PFC is $\pi$, as shown by the green curves. Hence, we can safely say that the small increment in $E^N_{c2-m}$ is due to the cutback in $E^N_{c2-c1}$ and $E^N_{c2-b}$.  

Interestingly, the robustness of the entanglement against temperature is also affected by the gain of the PFC, as shown in Fig. 5. In Fig. 5(a-c), we present contour plots of three bipartite entanglements $E^N_{c2-c1}$, $E^N_{c2-m}$, and $E^N_{c2-b}$ as a function of gain of the PFC and the temperature. It is vital to mention here that we have taken the optimal value of the detunings and phase of the PFC. It can be seen from Fig. 5(a-b) that the optimal value of entanglement $E^N_{c2-c1}$ ($E^N_{c2-m}$) can be observed for a range of parametric gain, i.e. $\xi/\omega_{b}\in[0.1 \longrightarrow 0.65]$ ($\xi/\omega_{b}\in[0.1 \longrightarrow 0.85]$). However, from Fig. 5(c), it is apparent that the optimal entanglement $E^N_{c2-b}$ is obtained for a wider range of parametric gain, i.e. $\xi/\omega_{b}>0.2$
Additionally, as the temperature increases, the range of parametric gain that can produce entanglement gradually contracts. Compared to $E^N_{c2-c1}$ and $E^N_{c2-m}$, $E^N_{c2-b}$ has a significantly larger robustness. However, both $E^N_{c2-m}$ and $E^N_{c2-b}$ cease to exist at a temperature of $185$ mK, while $E^N_{c2-c1}$ survives upto $140$mK.

Finally, let us talk about the asymmetric quantum steering that exists between the two modes of the subsystem. Figure 6 is a significant finding that demonstrates one-way steering between indirectly coupled modes, i.e. cavity-2-magnon and cavity-2-phonon. However, the steering between two cavity modes has not been observed. It can be seen that magnon and phonon mode maximally steer the cavity mode-2, when two cavity modes are tuned to the red (Stokes) sideband regime, while we observed no steering in swapped direction as shown in Fig. 6(a-b). In addition, the phonon mode steers the cavity mode-2 more strongly as compared to the magnon mode. Furthermore, we also observe small amount of steering $\xi_{b|c2}$ in blue detuned region. 
Notably, it is also possible to achieve the perfect transmission of the asymmetric quantum EPR steering in our system.
Cavity-2-magnon and cavity-2-phonon quantum steerings versus the gain of the frequency converter $\xi$ are plotted in fig. 6(c). One can see that the maximum of the Cavity-2-magnon steering $S_{m | c2 }$
emerges in the locality of $\xi=0.33\omega_{b}$, corresponding to the entanglement $E^N_{c2-m}$ as demonstrated in Fig. 6(c); however, no steering in swapped direction is observed. With enhancing $\xi$, the directional steering $S_{m | c2 }$ cease to exist and the asymmetric one-way steering between cavity-2 and phonon $S_{b | c2 }$ appears around $\xi=0.33\omega_{b}$, with the concealed steering between cavity-2 and magnon, and attain the optimum value around $\xi=0.45\omega_{b}$. 
That process indicates that we
achieve the perfect transfer of directional steering between the magnon-photon pair and the phonon-photon pair. Furthermore, one can notice that steering between the phonon-photon pair exists for a long range of parametric gain as compared to magnon-photon pair's steering.
One-way quantum steering bring forth one-side device independent quantum key distribution (QKD), which offers security protocol when only one party's measurement apparatus is trusted. Moreover, one-way quantum steering has been realized and experimentally implemented \cite{Hnd,SUN}.
\begin{figure}[b!]
\begin{center}
\includegraphics[width=1\columnwidth,height=3in]{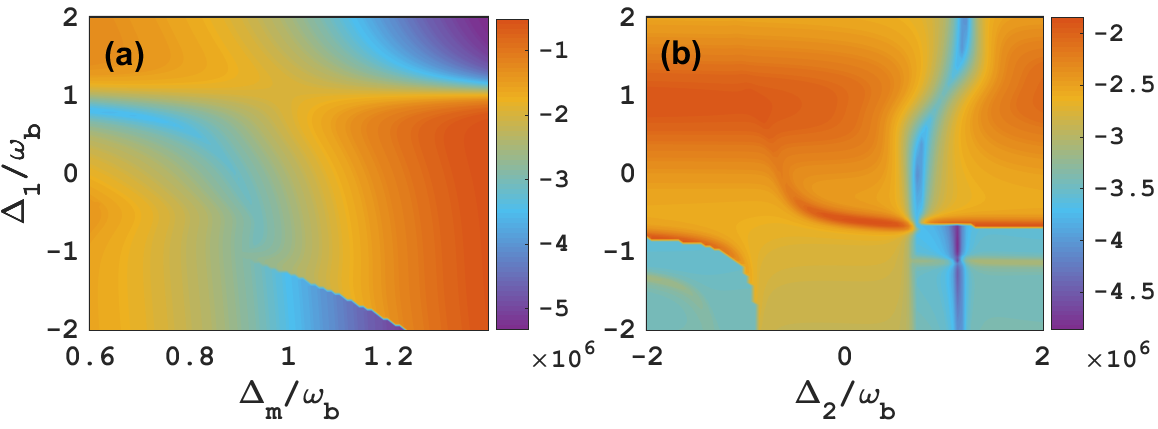}
\end{center}
\caption{Contour plot of the real part of the maximum eigenvalues of the drift
matrix versus $\Delta_{1}/\omega_{b}$ and (a) $\Delta_{m}/\omega_{b}$ (b) $\Delta_{2}/\omega_{b}$. The parameters are detailed in the text.}
\end{figure}
\section{CONCLUSION}
In this study, we focus on transfer of  entanglement and one-way quantum steering using a simple non-linear process of parametric frequency converter and have shown that parametric frequency converter plays an important role in the enhancement of two above quantum correlation. We considered a two-mode magnomechanical having a  yttrium-iron-garnet (YIG) sphere and a parametric converter. We exploited the magnetic dipole interactions between YIG sphere and cavity modes and magnetostrictive interaction between magnon-phonon modes. It has been shown that parametric converter acts as a source for enhanced entanglement among all bipartitions and asymmetrical quantum steering. Entanglement and quantum steering both are shown to be dependent on parametric gain and the associated phase factor $\phi$ of the PFC. It has been shown that one-way steering exists between indirectly coupled cavity modes, however, the steering between two cavity modes does not exist. In addition, it has been shown that the entanglement is more robust against thermal effects, with the addition of parametric frequency converter, as compared to the case of bare cavity. Throughout this work, we have assumed the adiabatic working regime and ignored the photon generation by other processes like retardation, Casimir and Doppler effects.
\section*{Appendix. Stability and Covariance matrix of the system}
The primary need for any system's efficacy is its stability. In other word, the stability condition limits the system's parameters to values at which the system remains stable. Hence, finding the parameters where stability occurs in the regime is therefore the primary objective. In this regard, the Routh-Hurwitz criterion must be met for the system to be considered stable \cite{RHCr,Shl3}, which says that the real part of all the eigenvalues of the drift matrix must be negative. Therefore, we first extract the eigenvalues from
the matrix $\mathcal{M}$ using the secular equation i.e., $|\mathcal{M}%
-\lambda _{\mathcal{M}}\mathbb{I}|=0$ (here $\Lambda_{M}$ are the eigenvalues of matrix M), and confirm the system's stability.
Based on our numerical calculation, we plot the maximum of the real part of the eigenvalues of the drift matrix $\mathcal{M}$, shown in Fig. 7(a-b), which show the stability of our system since the maximum of the real
parts of the eigenvalues are all negative. Moreover, Fig. 7(a–b) show that the system is more stable around $-\Delta_{1}=-\Delta_{1}=\Delta_{m}=\omega_{b}$, which endures our claim. In conclusion, the working regime falls within the stable regime since the experimental values used in the simulations meet the stability requirements of the Routh-Hurwitz criterion.

The covariance matrix $\mathcal{V}$ of the present two-mode cavity magnomechanical system is given by:%
\begin{equation}
\mathcal{V} =\left(
\begin{array}{cccc}
\Lambda _{b} & \Theta _{b.m} & \Theta _{b.c1} & \Theta _{b.c2} \\
\Theta _{b.m}^{T} & \Lambda _{m} & \Theta _{m.c1} & \Theta _{m.c2} \\
\Theta _{b.c1}^{T} & \Theta _{m.c1}^{T} & \Lambda _{c1} & \Theta _{c1.c2} \\
\Theta _{b.c1}^{T} & \Theta _{m.c2}^{T} & \Theta _{c1.c2}^{T} & \Lambda _{c2}%
\end{array}%
\right).  \label{Cv}
\end{equation}
Here, $\Lambda _{i}$ and $\Theta _{i.j}$ are $2\times 2$ matrices stand for the local and the inter-mode correlation properties of the phonon, magnon, and two cavity modes, for $i=b,m,c1,c2$, respectively. For example, $\Lambda _{m}$ ($\Theta _{m.c1}$) describe
the local (inter-mode) property of magnon mode (magnon-cavity-1 modes).
 
\section*{Acknowledgement}
We are thankful for the fruitful discussion with Prof. Dr. Shahid Qamar.

\section*{Declaration of interest}
The authors declare that none of the work described in this study could have been influenced by any known competing financial interests or personal relationships.

\section*{Data availability}
All numerical data that support the findings in this study is available
within the article.

\section*{ORCID iD}
Amjad Sohail https://orcid.org/0000-0001-8777-7928\\
Rizwan Ahmed https://orcid.org/0000-0001-8168-6777\\
Hazrat Ali https://orcid.org/0000-0003-1957-3629\\
M. C. de Oliveira https://orcid.org/0000-0003-2251-2632

\end{document}